\documentclass[a4paper,
               ]{jacow}
\usepackage{pdfpages,multirow,ragged2e} %
\makeatletter%
	\ifboolexpr{bool{xetex}}
	 {\renewcommand{\Gin@extensions}{.pdf,%
	                    .png,.jpg,.bmp,.pict,.tif,.psd,.mac,.sga,.tga,.gif,%
	                    .eps,.ps,%
	                    }}{}
\makeatother

\ifboolexpr{bool{xetex} or bool{luatex}} 
 {}                                      
 {\usepackage[utf8]{inputenc}}           

\usepackage[USenglish]{babel}

\ifboolexpr{bool{jacowbiblatex}}%
 {%
  \addbibresource{jacow-test.bib}
  \addbibresource{biblatex-examples.bib}
 }{}
\begin{document}

\title{High-precision RF voltage measurements using longitudinal phase-space tomography in CERN PSB and SPS}

\author{D. Quartullo\thanks{danilo.quartullo@cern.ch}, S. Albright, H. Damerau, G. Papotti, CERN, Geneva, Switzerland}
	
\maketitle

\begin{abstract}
   Precisely determining the gap voltage and phase in an RF cavity is essential for the calibration of the LLRF feedbacks. Following the conventional approach, measured RF power is converted into gap voltage, assuming a given shunt impedance. However, power and impedance evaluations can both have large uncertainties. Alternatively, the voltage can be obtained precisely with a technique based on longitudinal phase-space tomography. From a set of bunch profiles, tomography reconstructs the bunch distribution in the longitudinal phase-space. The quality of the reconstruction strongly depends on the RF voltage and therefore allows to derive its absolute value. In this paper we describe the tomography-based voltage measurements performed in the CERN PSB and SPS, where this method allowed to detect significant voltage errors for the main RF systems. After applying the correction factors in the LLRF, 1\% accuracies were reached. We report here also the remarkable results achieved by using this technique to calibrate the voltage of the SPS higher-harmonic cavities at 800 MHz, as well as their relative phases with respect to the 200 MHz cavities.
\end{abstract}

\section{INTRODUCTION}

The knowledge of the gap voltage in an RF cavity is important for several reasons.
Firstly, the RF voltage is necessary to calibrate the LLRF feedbacks, e.g.~for beam-loading compensation. Secondly, analytical beam dynamics studies can provide accurate estimates only if the accelerator parameters, including the RF voltage, are known with sufficient precision. Thirdly, an accurate evaluation of the RF power transmitted into the cavity can be obtained, allowing to determine the performance of the cavity or possible issues.  

However, measuring the RF voltage with conventional techniques can be difficult. One method consists of measuring indirectly the RF power driven into the cavity. This power is then converted into gap voltage assuming a certain shunt impedance. However, power and impedance measurements of high-frequency cavities can be affected by significant errors. In addition, the cavity model to convert the power into voltage introduces another source of uncertainty.

Beam-based techniques to measure the RF voltage analyze the longitudinal profiles of an oscillating bunch. The beam intensity is generally kept as low as possible, since collective effects induce an additional voltage in the cavity. The measurements are generally performed with beam-related loops of the LLRF system deactivated, to avoid damping the desired bunch oscillations and also to simplify the analysis. 

\begin{figure}[!h]
	\centering
	\includegraphics*[width=\columnwidth]{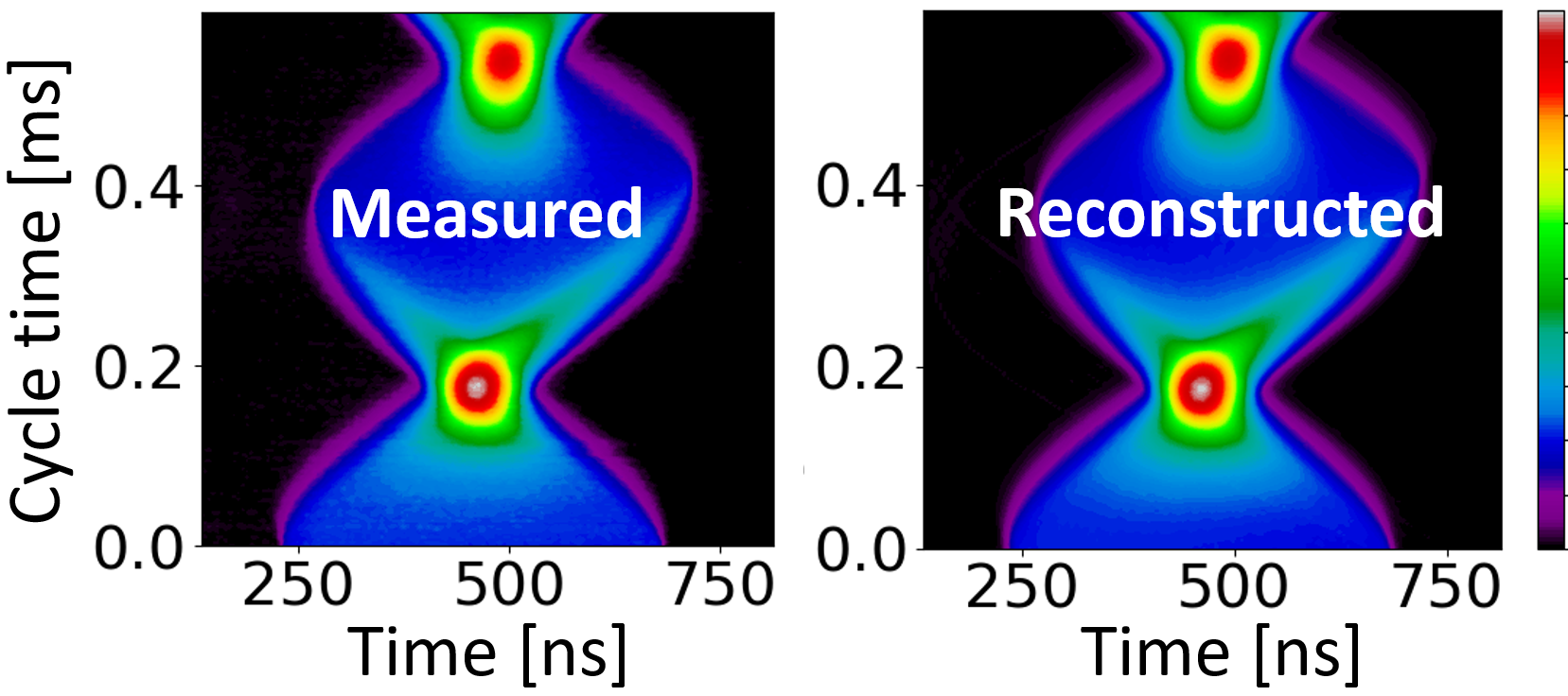}
	\caption{Evolution of bunch profiles measured at PSB flat-bottom, programmed $V_\textrm{p}=7.5$ kV for the cavity in sector S13 of ring 4 (left). Reconstructed bunch profiles (right). The optimal voltage and phase used for tomography derive from the voltage measurement shown in the top plot of Fig.~\ref{fig:example_1_MOD2}. }
	\label{fig:example_1_MOD1}
\end{figure}

\begin{figure}[!h]
	\centering
	\includegraphics*[width=\columnwidth]{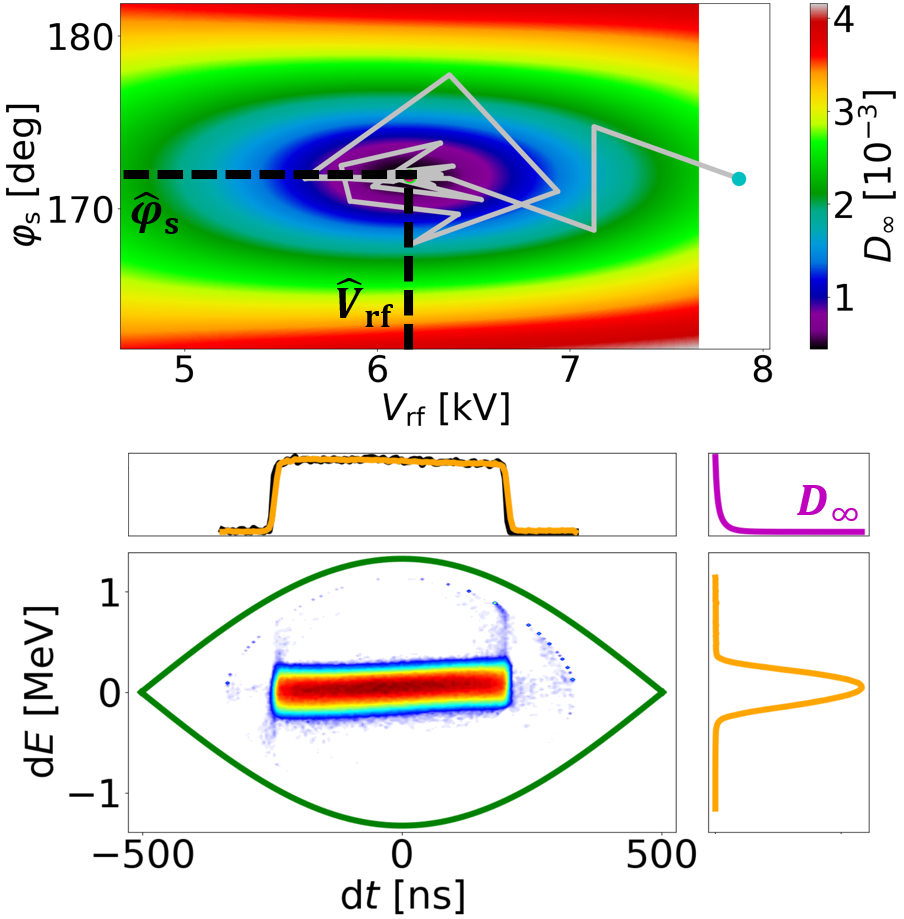}
	\caption{Tomography-based voltage measurement using as input the bunch profiles shown in the left plot of Fig.~\ref{fig:example_1_MOD1} (top). The voltage error is $-18\%$. The minimum discrepancy is found with a function-minimizer algorithm (grey curve), and alternatively by computing $D_{\infty}$ on a rectangular grid. Phase-space distribution corresponding to the first profile (bottom). The time and energy projections of the reconstructed distribution are plotted in orange. The reconstructed bunch profile overlaps the measured one, which is shown in black. The magenta curve represents the discrepancy as a function of the number of algorithm steps. The optimal voltage and phase are assumed for the tomographic reconstruction. }
	\label{fig:example_1_MOD2}
\end{figure}

The precision of synchrotron frequency measurements is limited, due to their dependence on the oscillation amplitude. A more evolved beam-based technique for RF voltage measurements uses the longitudinal phase-space reconstruction by tomography \cite{Hancock1, Hancock2, Hancock3, albright1, Grindheim}. Given a set of bunch profiles (Fig.~\ref{fig:example_1_MOD1}, left), tomography can reconstruct the bunch distribution in the longitudinal phase-space (Fig.~\ref{fig:example_1_MOD2}, bottom). Tomography is an iterative algorithm, and at each step the weighted difference (discrepancy, $D$) between reconstructed and input profiles usually decreases. The discrepancy is given by
\begin{equation}
    D = \sqrt{\frac{\sum_{i=1}^{N_{\textrm{fr}}} \sum_{j=1}^{N_{\textrm{bin}}} (\Delta P_{i,j})^2}{N_{\textrm{fr}}N_{\textrm{bin}}}},
    \label{eq:discr}
\end{equation}
where $N_{fr}$ is the number of profiles given as input to tomography, $N_{bin}$ is the number of bins per profile, and $\Delta P_{i,j}$ is the difference between the $i^{th}$ measured and reconstructed profiles at the $j^{th}$ bin. For a sufficiently large number of iterations (typically below 100), the discrepancy in Eq. (\ref{eq:discr}) converges to an asymptotic value $D_{\infty}$ (Fig.~\ref{fig:example_1_MOD2}, bottom), which indicates the quality of the phase-space reconstruction.  

The accuracy of a tomographic reconstruction strongly depends on the knowledge of the RF voltage, $\hat{V}_{\textrm{rf}}$ and of the phase of the bucket center, $\hat{\varphi}_{\textrm{s}}$. Hence, the asymptotic discrepancy, $D_{\infty}$ can be considered a function of the voltage and bucket position. The measurement of the RF voltage and phase using tomography consists in determining $\hat{V}_{\textrm{rf}}$ and $\hat{\varphi}_{\textrm{s}}$ which minimize the discrepancy function (Fig.~\ref{fig:example_1_MOD2}, top). This technique, already used at CERN in 1998 for the RF cavities of the PS accelerator \cite{Hancock5}, has been extensively applied since 2021 to measure with high precision the RF voltage of the PSB and SPS cavities \cite{quartullo1, quartullo2}. Moreover, tomography-based measurements allowed to determine the relative RF phases of the SPS cavities with remarkable accuracy.

\section{VOLTAGE MEASUREMENTS FOR THE PSB CAVITIES}

\subsection{Setup of the beam measurements}

The PSB has four superposed rings with three identical RF cavities each, in sections S5, S7 and S13. At injection energy ($E_{\textrm{\textrm{kin}}}=160$ MeV) and in each ring, the RF voltage of the main $h=1$ harmonic ($f_{\textrm{\textrm{rf}}}=1.0$ MHz) was measured separately in the three cavities in S5, S7 and S13. To check the linearity, the programmed voltages were 4~kV, 5~kV, 6~kV and 7.5~kV. For each combination of ring, cavity and programmed voltage, at least three cycles were measured to verify the reproducibility of the results.

The beam measurements started at injection and extended for the entire flat-bottom duration. The bunch injected from the Linac4 had a rectangular shape in the longitudinal phase space (Fig.~\ref{fig:example_1_MOD2}, bottom), therefore, being not matched to the RF bucket in the PSB, when rotating in phase space this distribution caused strong bunch length (quadrupole) oscillations (Fig.~\ref{fig:example_1_MOD1}, left). The bunch intensity was kept as low as possible, i.e. below $8\cdot 10^{10}$ ppb, to limit the influence of collective effects. The LLRF servo loops needed for beam-loading compensation in the cavities were kept active, however dedicated tests showed that their effect on the voltage measurements was negligible. Beam phase and radial loops were disabled during the measurements to remove their influence.

To study the dependence of the voltage errors on the RF frequency, measurements were also taken at extraction energy~($E_{\textrm{\textrm{kin}}}=2$ GeV, $f_{\textrm{\textrm{rf}}}=1.8$ MHz). The programmed voltages were the same as for the measurements at injection. To excite quadrupole bunch-oscillations, RF voltage jumps were programmed before the start of the flat-top. In addition, large dipole oscillations were induced by opening the phase loop before the start of the measurements. Longitudinal shaving of the bunch along the ramp allowed to obtain relatively low bunch intensities at flat-top (below $3\cdot 10^{10}$~ppb).

\subsection{Results}

\begin{figure}[h!]
	\centering
	\includegraphics*[width=\columnwidth]{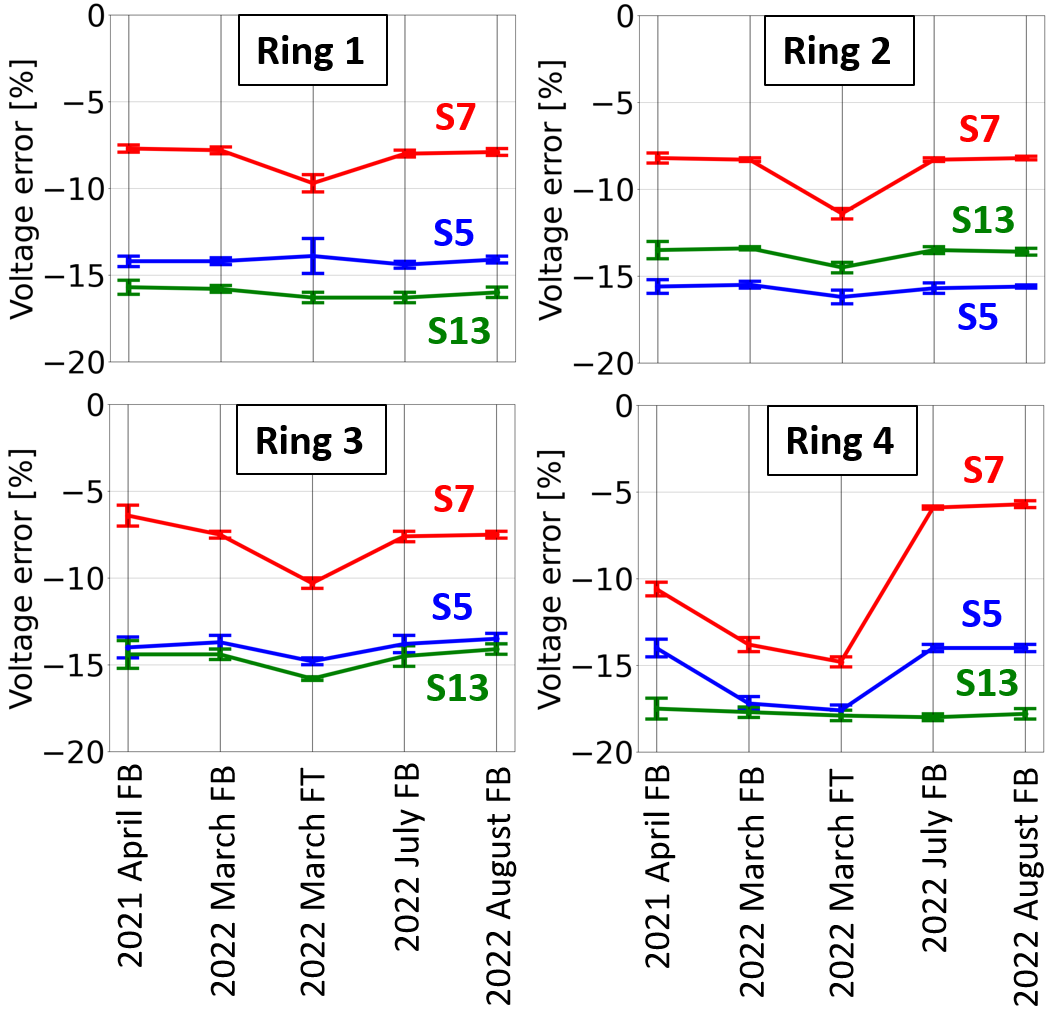}
	\caption{Evolution of the voltage errors over the years for the cavities in S5 (blue), S7 (red) and S13 (green), in all the four PSB rings, at flat-bottom (FB) or at flat-top (FT). The error bars associated to each set of measurements indicate the spread due to the different programmed voltages tested. At least three cycles were recorded for a given set of parameters, with cycle-to-cycle variations in voltage errors usually within $\pm 1\%$ (not shown in the plots).}
	\label{fig:PSB_summary}
\end{figure}

The first voltage measurements were performed in early 2021, and they were repeated four times in 2022 (Fig.~\ref{fig:PSB_summary}). The voltage errors summarized in Fig.~\ref{fig:PSB_summary} vary between $-5$\% and $-18$\%. The cavities in S7 provide the lowest voltage errors in all rings, whereas the cavities in S13 give the largest errors, except in ring 2. The variation of the voltage errors over time is in general small for a given cavity, and the largest errors are usually observed with the measurements at flat-top. The small error bars in Fig.~\ref{fig:PSB_summary} (spread for different RF voltages) indicate that the linearity is excellent, as relative errors are essentially independent of the absolute voltage.   

In 2022, validation tests were performed by applying in the LLRF the correction factors obtained from the voltage measurements. In each ring, the beam measurements were repeated using the three cavities together. As expected, the voltage errors decreased from $-13\%$ to essentially zero.

\subsection{Dependence of the voltage error on errors in design energy and transition gamma}

The accuracy of tomography-based voltage measurements depends on the knowledge of the Lorentz factor at transition, $\gamma_\textrm{t}$ and of the dipole magnetic field acting on the bunch, $B$. To quantify these dependencies, longitudinal beam dynamics simulations with collective effects and assuming measured beam parameters were performed with the CERN BLonD code \cite{blondCode}. In the simulations, an RF voltage of 7.5 kV was assumed, whereas the transition gamma and the design magnetic field were set to expected values, i.e. $\gamma_\textrm{t}=4.05$, $B=0.232$ T at injection energy, $B=1.13$ T at flat-top. The simulated bunch profiles were used as input for the virtual voltage measurements, and errors in design magnetic field and transition gamma were intentionally introduced to determine their impact on the reconstructed RF voltage.  

\begin{figure}[h!]
	\centering
	\includegraphics*[width=\columnwidth]{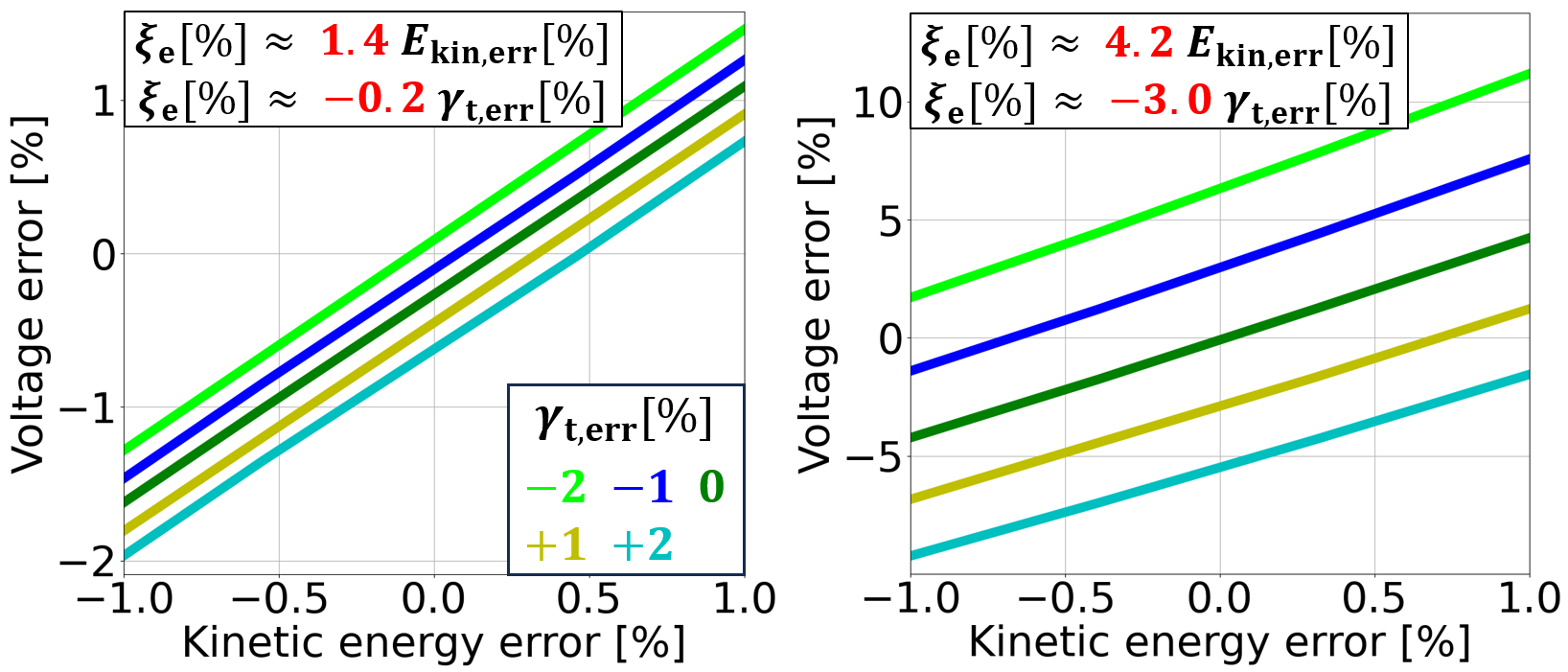}
	\caption{Voltage errors as a function of the error in the design kinetic energy, and also as a function of the transition gamma error, which is reported in the bottom-right box for each curve (left). The voltage errors are obtained by performing virtual voltage measurements at PSB flat-bottom, with simulated profiles as input. In the BLonD simulation, the assumed and expected kinetic energy and transition gamma are respectively 161 MeV and 4.05. The dependencies of the voltage errors are reported in the top-left box. Voltage errors as a function of the errors in kinetic energy and transition gamma at flat-top, assuming in simulation a kinetic energy of 2.02 GeV and a transition gamma of again 4.05 (right).}
	\label{fig:PSB_dependence}
\end{figure}

Figure \ref{fig:PSB_dependence} (left) shows the voltage-error inaccuracies at flat-bottom, as a function of relative errors in transition gamma, $\gamma_{\textrm{t},\textrm{err}}$ and in design kinetic energy, $E_{\textrm{kin},\textrm{err}}$. In relative units and in first approximation, the voltage errors increase linearly as a function of the kinetic energy error with a proportionality factor of 1.4, and they also increase linearly as a function of the transition gamma error with a factor of 0.2. A similar analysis at flat-top (Fig.~\ref{fig:PSB_dependence}, right) reveals larger factors of 4.2 and 3.0, respectively for the errors in kinetic energy and transition gamma. This indicates that the voltage measurements should be performed at flat-bottom to obtain better accuracies.

In the PSB, the expected uncertainty in the design magnetic field is about $0.1$ mT. Since the contribution of the dipole-magnets' bending radius to the uncertainty is considered to be negligible, the kinetic energy is expected to be accurate within $0.1\%$ and $0.02\%$, respectively at flat-bottom and at extraction energy. The uncertainty in the transition gamma is less well known. Note that a rounded value of $\gamma_{\textrm{t}}=4.1$ has been assumed for the voltage measurements reported in Fig.~\ref{fig:PSB_summary}. If the actual value of the transition gamma is indeed 4.05, then an error of $1\%$ has been introduced. Taking into account the proportionality factors reported above, this should lead to voltage-error underestimations of $1\%$ and $3\%$, respectively at injection and extraction energies (yellow lines in Fig.~\ref{fig:PSB_dependence}). In such a case, the results shown in Fig.~\ref{fig:PSB_summary} should be adjusted accordingly. This would lead to an even better voltage-error agreements at injection and extraction energies for the cavities in S7.

\section{VOLTAGE MEASUREMENTS FOR THE SPS FUNDAMENTAL CAVITIES}

\subsection{Setup of the beam measurements}

The SPS has six accelerating traveling-wave RF cavities, which are enumerated from 1 to 6 in the order of the beam direction. Each of these cavities operates at 200 MHz and is composed of either 33 (C1, C2, C4, C5) or 44 (C3, C6) cells. Beam measurements were performed at injection (26~GeV/c) and flat-top momentum (200 GeV/c) for each of the six cavities individually. The natural phase and energy mismatch of the bunch injected from the PS, combined with an intentional phase shift programmed in the SPS, excited large bunch oscillations at flat-bottom. For the measurements at 200 GeV/c, an RF phase jump before the start of the flat-top led to strong dipole oscillations of the beam. The programmed voltages were 950 kV and 1.6 MV, respectively for each 33-cell and 44-cell cavity. 

A dedicated bunch with low intensity ($4\cdot 10^9$ ppb) and short bunch length (1.2 ns) was injected into the SPS. The beam loops were turned off, but it was necessary to keep the One Turn Delay Feedback~(OTFB) for beam-loading compensation enabled, since its set point defined the voltage reference. Nevertheless, virtual voltage measurements with simulated profiles showed that the impact of the OTFB on the voltage errors is negligible at the given bunch intensity.

\subsection{Results}

\begin{figure}[!h]
	\centering
	\includegraphics*[width=\columnwidth]{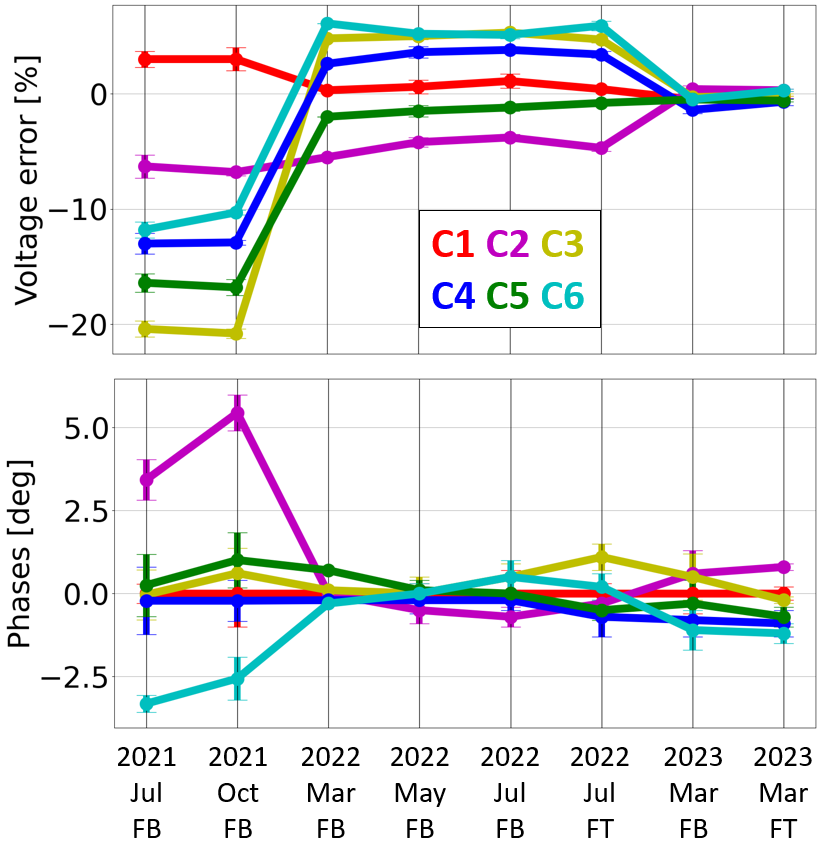}
	\caption{Evolution of the voltage errors~(top) and relative phases (bottom) for the six SPS cavities. The beam measurements performed at flat-bottom (26 GeV/c) or at flat-top (200~GeV/c) are identified respectively with `FB' and `FT'. In the bottom plot, the RF phase for the C1 cavity is arbitrarily set to zero as reference. For each set of measurements, the error bar indicates the spread due to cycle-to-cycle variation.}
	\label{fig:SPS_comparison_200MHz}
\end{figure}

The first voltage measurements for the 200 MHz cavities were performed in 2021, and errors up to 20\% were found (Fig.~\ref{fig:SPS_comparison_200MHz}, top). A careful check revealed the presence of inaccurate settings for the variable attenuators used to calibrate the voltage in the cavities. These settings were due to errors in power estimates. Refined electrical voltage calibrations were performed in early 2022. Afterwards, tomography-based voltage measurements indicated residual voltage errors within $6\%$, confirming that the electrical measurement campaign was successful. In early 2023, the correction factors obtained from the tomography-based voltage measurements were applied in the LLRF system. The voltage errors measured with beam in 2023 were within just $1\%$ with respect to the programmed voltage, which is an exceptional result. It is worth highlighting the excellent reproducibility within the same year of the voltage errors, as well as the small cycle-to-cycle spread.

The phase alignment among the six cavities was performed in 2021 and 2022. Tomography-based phase measurements revealed cavity misalignments of $8^{\circ}$ and $2^{\circ}$, respectively in 2021 and 2022 (Fig.~\ref{fig:SPS_comparison_200MHz}, bottom). No phase alignment was needed in 2023, since the cavities were still aligned within $2^{\circ}$. 

\subsection{Dependence of the voltage error on errors in design momentum and transition gamma}

As described above for the PSB case, the voltage-error inaccuracies as a function of errors in design momentum and transition gamma were evaluated also in the SPS by performing virtual voltage measurements with simulated profiles. At flat-bottom, the approximately linear dependencies have proportionality factors of 0.4 and 3.5, respectively for relative errors in momentum and transition gamma. The dependencies are roughly linear also at 200 GeV/c, with a factor of 1.0 for errors in momentum, and of 2.0 for relative errors in transition gamma. Therefore, voltage measurements should be performed at flat-bottom or at 200 GeV/c, depending on the expected uncertainties.  

In measurements, the design magnetic field is expected to be accurate within $0.3$ mT. The uncertainty in momentum is of $0.3\%$ and $0.03\%$, respectively at injection energy and at flat-top. The uncertainty in transition gamma is not well known, although it is expected to be lower than 1\%. Indeed, taking into account the proportionality factors reported above, an error of 1\% in transition gamma would lead to voltage errors at flat-top consistently smaller by $1.5\%$ than the ones at flat-bottom, but this does not occur (Fig.~\ref{fig:SPS_comparison_200MHz}, top). Assuming for instance an error of 0.5\% in transition gamma, the voltage errors would be accurate within just 2\% and 1\%, respectively at flat-bottom and at 200 GeV/c.

\section{VOLTAGE MEASUREMENTS FOR THE SPS HIGHER-HARMONIC CAVITIES}

\subsection{Voltage measurements in a double-RF system}

In the SPS, an RF system operating at 800 MHz is used together with the accelerating one to increase the stability of the accelerated bunches. The relative phase between the two RF systems is chosen so that the bunch length decreases compared to the single-RF case (bunch shortening mode). The two 800 MHz traveling-wave cavities $\textrm{C}1_{800}$ and $\textrm{C}2_{800}$, enumerated with 1 and 2 in the order of the beam direction, are composed of 37 cells each. Tomography-based measurements in a double-RF system allowed to determine the voltage, $\hat{V}_{\textrm{rf},2}$ of each 800 MHz cavity, as well as the relative phase, $\hat{\varphi}_{12}$ between the 200~MHz and 800 MHz RF systems. 

In a double-RF voltage measurement, the discrepancy depends on $V_{\textrm{rf},1}$, $\varphi_{\textrm{s}}$, $V_{\textrm{rf},2}$ and $\varphi_{12}$. Therefore, the goal is to minimize it as a function of these four parameters. This can be computationally challenging, also because the algorithm for function minimization can get stuck in one of the many local minima of the four-dimensional parameter space. Alternatively, a given four-parameter voltage measurement can be replaced by a sequence of two-parameter voltage measurements. The first of these takes as input profiles measured in single RF and provides as output $\hat{V}_{\textrm{rf},1}$ and $\hat{\varphi}_{\textrm{s}}$. These optimal parameters are needed as an input for the second two-parameter voltage measurement, which provides $\hat{V}_{\textrm{rf},2}$ and $\hat{\varphi}_{12}$ starting from profiles recorded in double RF.   

\subsection{Results}

\begin{figure}[!h]
	\centering
	\includegraphics*[width=\columnwidth]{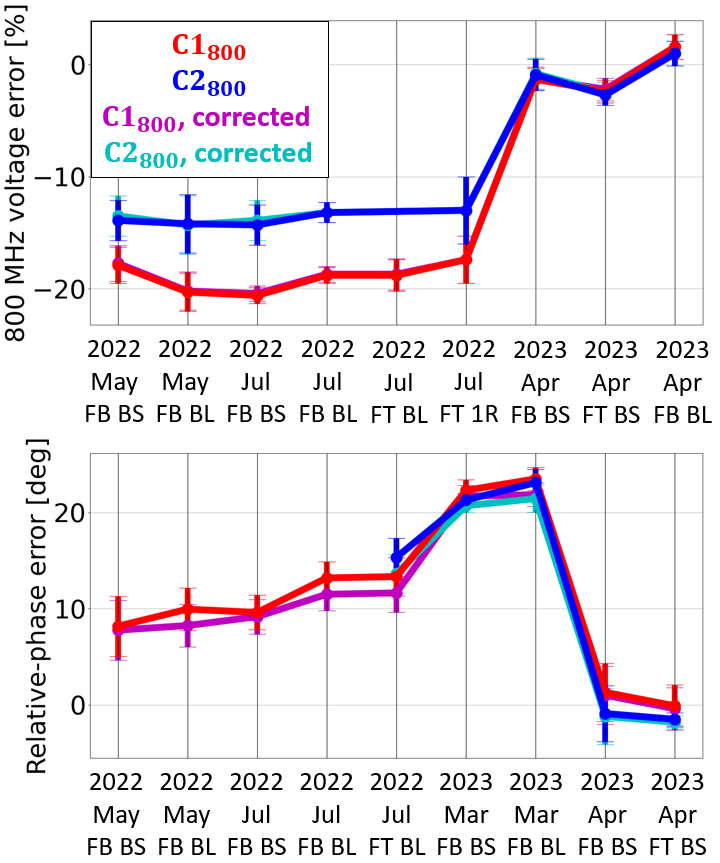}
	\caption{Evolution of the voltage errors for the SPS 800~MHz RF cavities (top) and of the relative phase errors for the 200 MHz and 800 MHz RF systems (bottom). The measurements at flat-bottom (26 GeV/c) or at flat-top (200~GeV/c) are identified respectively with `FB' and `FT'. The measurements performed in a double-RF system in bunch shortening or lengthening mode are respectively denoted with `BS' and `BL'. The measurements in the 800~MHz single-RF system are identified with `1R'. The red and blue curves represent the uncorrected results, whereas the other two curves indicate the voltage errors after removing the collective-effects contribution, which is estimated by performing virtual voltage measurements. The error bars indicate the spreads due to cycle-to-cycle variation.}
	\label{fig:SPS_comparison_800MHz}
\end{figure}

\begin{figure}[!h]
	\centering
	\includegraphics*[width=\columnwidth]{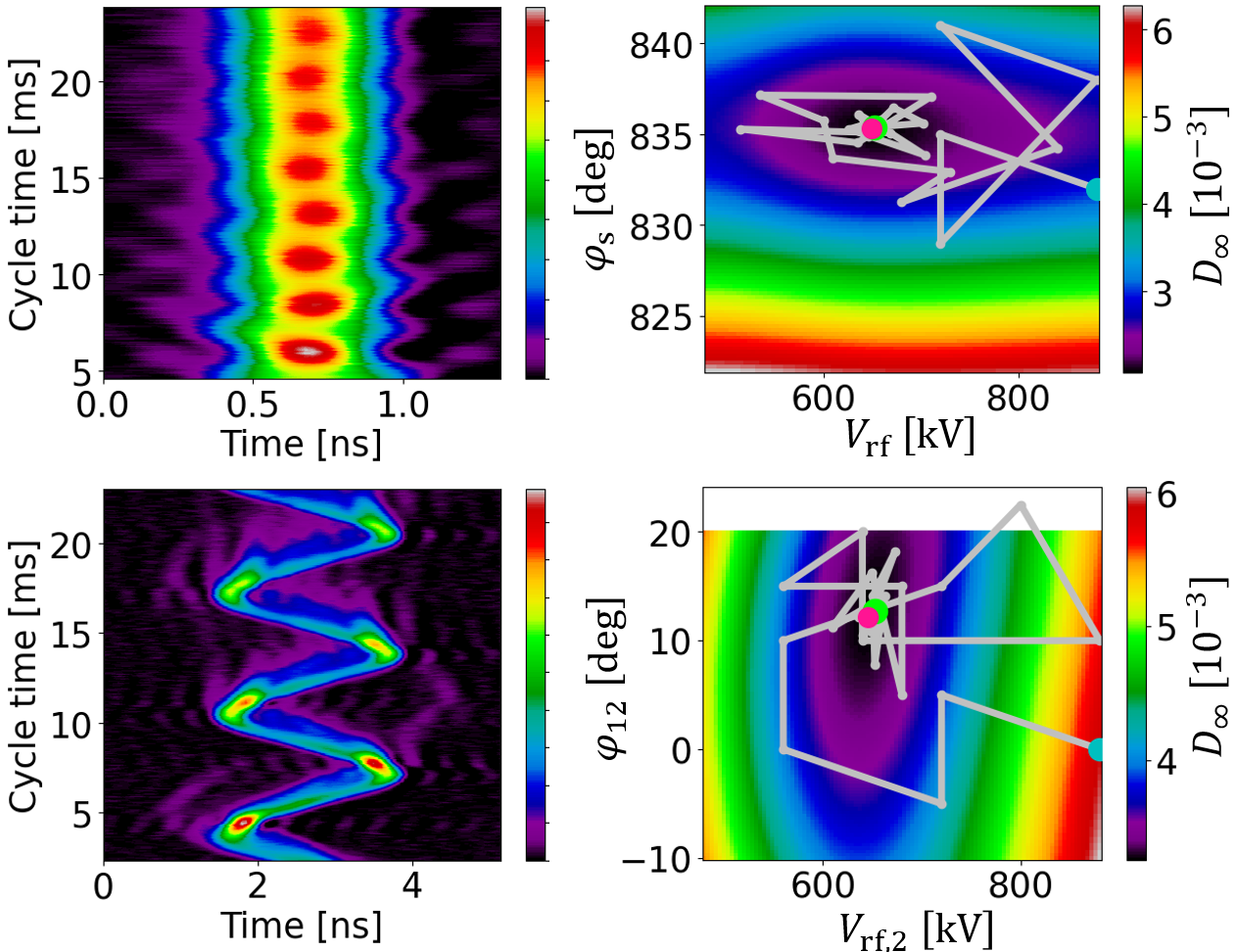}
	\caption{Evolution of bunch profiles measured in 2022 at 200 GeV/c, with $\textrm{C}1_{800}$ alone (top left) as well as with the pair of RF cavities C1 and $\textrm{C}1_{800}$ (bottom left). The programmed voltage for $\textrm{C}1_{800}$ is $V_\textrm{p} = 800$ kV. The voltage measurements (right) use as input the bunch profiles shown in the corresponding left plots. The voltage error for $\textrm{C}1_{800}$ is $-18\%$ in both cases.}
	\label{fig:example_2}
\end{figure}

Voltage measurements in the double-RF system were performed for the first time at flat-bottom in 2022, and they revealed voltage errors of $18\%$ and $14\%$, respectively for $\textrm{C}1_{800}$ and $\textrm{C}2_{800}$ (Fig.~\ref{fig:SPS_comparison_800MHz}, top). These errors were confirmed by performing beam measurements also at 200~GeV/c, in single and double-RF system (Fig.~\ref{fig:example_2}). In early 2023, the correction factors obtained from the tomography-based voltage measurements were applied in the LLRF system. Beam measurements were then repeated, and they revealed voltage errors within just $2\%$ (Fig.~\ref{fig:SPS_comparison_800MHz}, top), which is a remarkable result. The contribution of collective effects on the voltage errors was again estimated by performing virtual voltage measurements using simulated profiles. As Fig.~\ref{fig:SPS_comparison_800MHz} (top) shows, this contribution can be neglected.

In 2022, the calibration of the relative phase between the 200 MHz and 800 MHz RF systems was performed by evaluating the bunch-profile tilt as a function of the programmed relative phase. The more precise tomography-based measurements revealed relative phase errors between $8^{\circ}$ and $14^{\circ}$ for the pair of RF cavities C1 and $\textrm{C}1_{800}$ (Fig.~\ref{fig:SPS_comparison_800MHz}, bottom). In early 2023, tomography-based measurements revealed relative phase errors of $23^{\circ}$ for each 800 MHz cavity, therefore relative phase corrections of $23^{\circ}$ were applied in the LLRF system. The beam-based measurements were repeated, and relative phase errors within just $2^{\circ}$ were found, which demonstrates the precision of the technique. Figure \ref{fig:SPS_comparison_800MHz} (bottom) also shows that the estimated contribution of collective effects on the phase errors is small.

\section{CONCLUSION}

Determining the exact gap voltage in an RF cavity can be difficult using the conventional approaches. A more accurate technique, based on beam-profile measurements and tomographic reconstructions, has been used at CERN to measure the RF voltages and phases of all the PSB and SPS RF cavities. 

Tomography-based voltage measurements in the PSB initially revealed voltage errors up to $18\%$. Applying the correction factors in the LLRF system, the voltage errors are reduced to less than $1\%$. In the SPS, first beam measurements for the 200~MHz cavities revealed voltage errors up to $21\%$. Refined electrical voltage calibrations were able to bring these errors down to $6\%$. The corrections from tomography-based voltage measurements were applied in the LLRF, and the voltage errors reduced to less than $1\%$. A very good phase alignment of the 200 MHz cavities could also be achieved. First beam measurements in a double-RF system revealed voltage errors up to $18\%$ for the 800~MHz cavities, as well as relative-phase errors of about $10^{\circ}$. In 2023, after having applied the corrections, voltage and relative phase errors were below $2\%$ and $2^{\circ}$, respectively.

Given their precision, tomography-based voltage measurements are expected to become the baseline choice for voltage and phase measurements in all the synchrotrons at CERN. 

\section{ACKNOWLEDGEMENTS}

We thank our colleagues of the PSB and SPS LLRF teams for the support provided. We are grateful to A. Lasheen for his useful suggestions on tomography, and to I. Karpov for his help during beam measurements in the SPS. 

%
%
\ifboolexpr{bool{jacowbiblatex}}%
	{\printbibliography}%
	{%
	
	
} 
%
%


\end{document}